\documentclass[10pt,conference]{IEEEtran}
\IEEEoverridecommandlockouts

\def\BibTeX{{\rm B\kern-.05em{\sc i\kern-.025em b}\kern-.08em
    T\kern-.1667em\lower.7ex\hbox{E}\kern-.125emX}}

\usepackage{amsthm}
\usepackage{graphicx}
\usepackage{caption}
\usepackage{subcaption}
\usepackage{tabularx}
\usepackage[export]{adjustbox}
\usepackage{balance}

\usepackage{url}
\usepackage{booktabs}
\usepackage{array,multirow}
\usepackage{rotating}
\usepackage{float}
\usepackage{xcolor}
\usepackage{dblfloatfix}    
\usepackage{amssymb}

\usepackage{color}
\usepackage{tcolorbox}

\theoremstyle{plain}

\newtheorem*{definition*}{Definition}

\begin{document}

\title{Keeping Mutation Test Suites Consistent and Relevant with Long-Standing Mutants}

 \author{ 
\IEEEauthorblockN{
Milos Ojdanic}
\IEEEauthorblockA{\textit{University of Luxembourg} \\
 \textit{milos.ojdanic@uni.lu}}
 \and
 \IEEEauthorblockN{
 Mike Papadakis}
 \IEEEauthorblockA{\textit{University of Luxembourg} \\
 \textit{michail.papadakis@uni.lu}}
 \and
 \IEEEauthorblockN{
 Mark Harman}
 \IEEEauthorblockA{\textit{UCL and Meta platforms Inc.} \\
 \textit{mark.harman@ucl.ac.uk}}
}




\maketitle

\begin{abstract}
Mutation testing has been demonstrated to be one of the most powerful fault-revealing tools in the tester's tool kit.
Much previous work implicitly assumed it to be sufficient to re-compute mutant suites per release.
Sadly, this makes mutation results inconsistent; mutant scores from each release cannot be directly compared, making it harder to measure test improvement.
Furthermore, regular code change means that a mutant suite's relevance will naturally degrade over time.
We measure this degradation in relevance for 143,500  mutants in 4 non-trivial systems finding that, on overage, 52\% degrade.
We introduce a mutant brittleness measure and use it to audit software systems and their mutation suites.
We also demonstrate how consistent-by-construction long-standing mutant suites can be identified with a  10x improvement in mutant relevance over an arbitrary test suite.
Our results indicate that the research community should avoid the re-computation of mutant suites and focus, instead, on long-standing mutants, thereby improving the consistency and relevance of mutation testing.

\end{abstract}


\section{Introduction}

\begin{figure*}[h]
    \centering
    \vspace{-1em}
    \includegraphics[width=\textwidth]{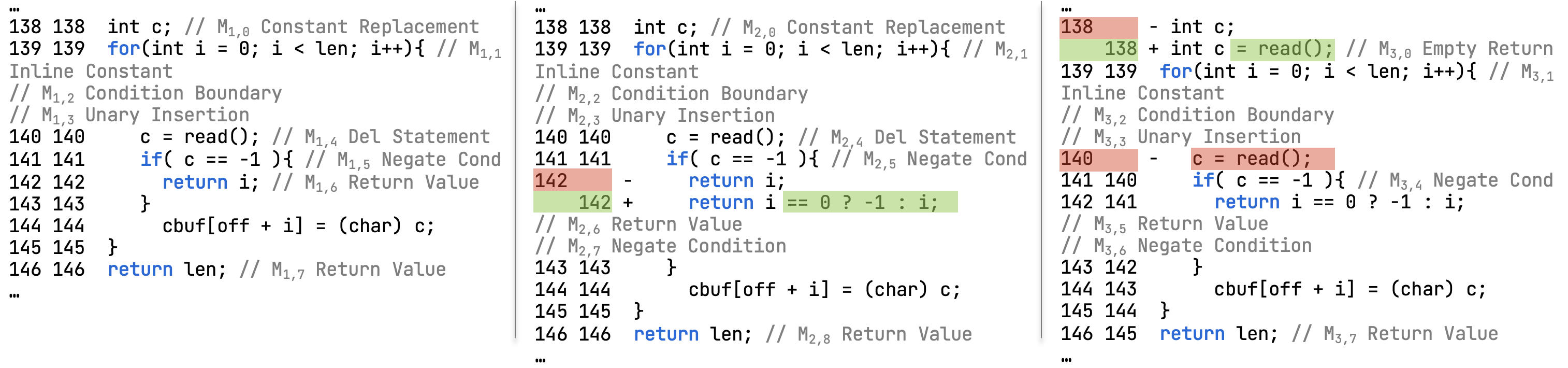}
    \vspace{-2em}
    \caption{\textbf{Example of mutants standing through 3 chronological sequences of code versions. The example code snippet comes from
Apache commons-io project, while method read() is excerpted from the BoundedReader.java (versions around 81210eb). The green and red rectangles represent associated commit changes. While java comments (//) describe the set of mutants $M_{i,j}$, where $i$ is the observed program version, and $j$ is a mutant ID.}}
    \label{fig:Long_standing_mutants_example}
    \vspace{-1em}
\end{figure*}


Mutation Testing has been  demonstrated to be one of the Software Testing community's most effective software testing techniques \cite{Chekam2017}.
It  seeds artificial faults - also known as mutants.
When a test distinguishes the behaviour of the mutant from the original, the mutant is said to be `killed'.
Test effectiveness is captured by the mutation score (the proportion of mutants killed) \cite{Papadakis2019,yjmh:analysis,offutselection93}.

Initial barriers  to adoption concerned the computational cost of the approach~\cite{jia2009higher, RegressionMutation2012}.
However, more recent techniques have tackled the  mutation cost problem using
intelligent mutant selection \cite{jia2009higher,RegressionMutation2012},
higher order mutation \cite{yjmh:homt,omar:subtle},
commit awareness \cite{OjdanicMLCVP22,ojdanic2022}
and
mutant subsumption \cite{9677967,Mark14,Ammann2014}, thereby removing these cost-based barriers.

We argue that there is a remaining barrier to uptake: mutant consistency.
We need a {\em consistent} set of mutants for a project so that test effectiveness can be consistently tracked against a common baseline over a series of project releases.
Sadly, almost all existing research on mutation testing assumes that a fresh set of mutants is created for each release of the system \cite{Papadakis2019,yjmh:analysis}.
An alternative would be to fix a set of mutants as a baseline and use this to measure ongoing test effectiveness evolution.
However, as the results of this paper show, such a fixed mutation set will quickly degrade in its relevance.

We  introduce a mutation test brittleness metric, which can be used to assess a mutation suite, and software project, in terms of the rate at which mutant relevance decays over a series of releases.
Our results demonstrate that mutants have diverse life spans across program versions. 
We show that identifying a high-quality suite of {\em long-standing} mutants allows us to maintain mutant relevance over a series of releases: a long-standing mutant suite provides test effectiveness relevance for at least 10x longer than a randomly selected suite. 
In order to increase consistency and relevance of mutation testing, we conclude that the research community should focus on  long-standing mutants, their applications, the opportunities they open, and the remaining open questions.

Specifically, this paper's primary contributions are:

\begin{enumerate}
    \item The introduction of long-standing mutants, as an important category warranting further study.
    \item The introduction of metrics for assessing mutant brittleness, and visualisations of how this metric varies for a given project over time.
    \item An empirical study of long-standing mutants, based on four non-trivial systems and 143,500 mutants.
    \item The key motivating finding that long-standing mutant suites enjoy an order of magnitude longer relevance than a randomly selected suite, over the four systems studied. 
    \item An important `special relationship' between long-standing and subsuming mutants: mutants that are subsuming in one version have a high probability of subsuming in the following versions. 
    This relationship is  important because it opens optimistic prospects for mutants' ability to maintain consistency, relevance, effectiveness {\em and} subsumption over a series of releases.

\end{enumerate}

\section{Background and Related Work}

\subsubsection*{\textbf{Commit-Relevant Mutants}}  Applying traditional mutation testing in CI processes is unpractical due to its cost. Meanwhile, Commit-Aware Mutation Testing scales and defines commit-relevant mutants as a set of mutants affected by the changed program behaviour that serve as commit-relevant test requirements to guide test assessment by aiming at the changed program functionality \cite{ojdanic2022, Cachia2013}. 
More recently, learning-based approaches \cite{MuDelta21} emerged capable of learning the commit-relevant mutants defined by a regression change in commit time, thus showing potential and opening a direction towards learning mutant's behaviour in evolving context. However - it is necessary to realise that to improve the testing process continuously and thus quantify overall testing quality - it would require applying the technique after each program change cycle, 
to not indebt and lose test requirements. The merit of reappearing mature mutants is that they preserve test requirements and thus complement commit-relevant mutants. In particular, the long-standing mutants promise to keep overlooked testing requirements from oblivion and provide test assessment for a prolonged time. 

\begin{table}[ht]
\caption{\small{Observed files through projects evolution}}
\vspace{-1em}
\resizebox{\columnwidth}{!}{
\centering
\begin{tabular}{l|r|r|r}
\toprule
\textbf{Observed Files} & \textbf{Time points} & \textbf{Mutants} &  \textbf{Commons Project} \\
\midrule
CSVParser & 31 & 8757 & csv \\
CSVRecord & 16 & 2656 & csv \\
Lexer & 21 & 10688 & csv \\
CSVLexer & 17 & 11208 & csv \\ 
CSVFormat & 47 & 52432 & csv \\
CSVPrinter & 21 & 20382 & csv \\ 
\midrule
IterableUtils & 10 & 6441 & collections  \\
\midrule
CharSequenceUtils & 10 & 6802 & lang \\
\midrule
WordUtils & 15 & 24128 & text \\
\bottomrule
\end{tabular}
}
\vspace{-0.1em}
\label{tab:Observed_history_of_files}
\end{table}

\subsubsection*{\textbf{Subsuming Mutants}} In an attempt to further scale and make test assessment affordable, many recent studies (consult the survey by Papadakis et al. \cite{Papadakis2019, PapadakisCT18}) consider subsuming mutants to reduce the number of mutants required to measure test adequacy \cite{Papadakis2019}. Indeed traditional mutation opperators introduce many trivial, duplicated, equivalent and redundant mutants \cite{PapadakisJHT15, PapadakisCT18}. Specifically, a subsumption relationship between mutants emerges from mutant behaviours, thus suggesting that the majority of the mutants fall into the redundancy basket since distinguishing those subsuming mutants will lead to the identification of all other mutants \cite{KintisPM10}. 

More formally, given a finite set of mutants \textit{M} and a finite set of tests \textit{T}, mutant \textit{$m_i$} is said to dynamically subsume mutant \textit{$m_j$} if some test in \textit{T} kills \textit{$m_i$} and every test in \textit{T} that kills \textit{$m_i$} also kills \textit{$m_j$} \cite{Ammann2014}. Calculating test effectiveness over subsuming mutants offers a much better test effectiveness indicator than the traditional mutation score since subsuming mutants have an almost linear relationship between number of test requ

Although the evidence is strong and the benefits are multi-fold, calculating subsumption in real time requires knowledge of mutant's behaviour, usually represented through test execution, which is unpractical in real-time. Recently few approaches have used learning-based methods to target subsuming mutants with a certain level of confidence, considering their location and properties \cite{9677967}. In this paper we reuse the guarantee of the quality of test assessment when the mutants are {\em also} long-standing.

\section{Long-Standing Mutants}

\subsection{Motivating Example}

A key challenge in the current state of mutation regression testing is a sequential version to version execution. Suppose we keep a record of generated mutants to the same element on which the mutant is generated and version them from one version to another.
Figure \ref{fig:Long_standing_mutants_example}, depicts a chronological sequence of 3 different versions of method \texttt{read()} extracted from Apache Commons-io project. 
Following the evolution of the code we can observe the evolution of the mutant set. We notice that most mutants reside in the same place across the versions. 
When a mutant location is unchanged from version to version, we consider that it stands through time. While if a mutant does not occur in next version, we consider it to stop standing, e.g., $M_{2,4}$ does not exist as $M_{3,4}$ due to deletion changes. 
From the example, we can observe when a system reaches maturity, as in the case of commons-io, the majority of the changes do not touch the core logic and most mutants $M_{1,n}$ are long-standing (still exist as $M_{3,n}$) precisely six out of eight.
In particular, this indicates potential reuse of past subsumption knowledge concerning selection priority (among other things), avoiding redundancy, addressing technical testing debt and aspiring towards test completeness of mature code components. 


\vspace{-0.5em}
\subsection{Implementation Details - Mapping}

\begin{figure}[t]
    \centering
    \includegraphics[width=\linewidth]{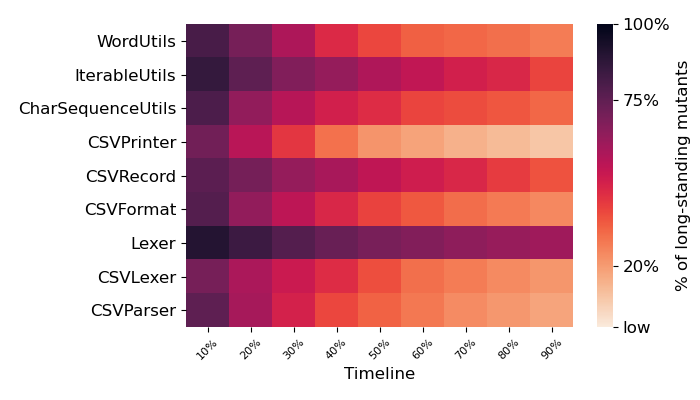}
    \vspace{-2em}
    \caption{\textbf{Mutant sets for different observed files and their corresponding studied history (timeline). Each cell in the heat map represents a normalised proportion of a subject history length, and the corresponding colour represents a percentage of mutants from the initial set over time through versions. The results show that mutants have diverse longevity, with brittleness, on average, of 52\%.}}
    \label{fig:Distribution_of_long_living_mutants}
    \vspace{-2em}
\end{figure}

To reuse mutants and follow their standing (w.r.t. how long a mutant `stands' in one location without being altered), in this study, we map mutants considering changed lines and the context of change from git diff tool \cite{GitDiff}. Our scripts take two program versions and map code statements from version to version. Figure \ref{fig:Long_standing_mutants_example} indicates information of line numbers shift from version to version. \textit{Note that for the purpose of this study, the history length of a mutant is computed from the first studied version till the last chronologically observed version}.

We start our analysis by extracting files with the most extended history of change from the open-source mature Apache Commons projects. Then, we use the state-of-the-art PIT mutation testing tool \cite{ColesLHPV16} to generate mutants per each changed (committed) file, followed by the execution of tests and generation of the killing matrix. Next to the killing matrix, for each changed file, we keep metadata (info. about hunks, timestamps, mutants bytecode index, location etc.) Using extracted information, we create a regression history for each file, making a file-specific historical timeline. In the  timeline of each file, a time-point represents a commit that introduces changes to the file. While for each time-point, we calculate subsuming mutants (\textit{reminder:} the set of mutants, when distinguished, distinguish all other mutants).

Besides the set of mutants, each time-point contains information about mapping changes to the consecutive time-point. Hence, long-standing mutation metric is a function \textit{$F(M_t, $change$\_$map$) = M_t'$}. Where \textit{$M_t$} is a mutant from time \textit{t}, and \textit{$change$\_$map$} is a map containing information about code transition from time \textit{t $\rightarrow $t'}, while \textit{$M_t'$} is the mutant at time \textit{t'}.
Therefore long-standing mutants definition is: 


\begin{definition*}
Given \textit{n} and \textit{m} as timepoints of the first and last versions under the study of program \textit{P}, where \textit{n $>$ m}. A mutant \textit{M} is said to be long-standing if it exists on the same code element \textit{E} throughout consecutive versions \textit{$P^n$}, \textit{$P^{n+1}$}, ... ,\textit{$P^m$} until the point when the mutant \textit{M} due to a committed program change does not appear on the code element \textit{E} of a program version \textit{$P^{m+1}$}.
\end{definition*}


Given that  mutants can `stand' for several versions or just a few, we argue that the rate at which a mutation suite and mutant relevance decays over a series of releases suggests a mutation test brittleness. Knowing the degree to which mutants hold high-quality tests for a series of releases helps provide more prolonged test effectiveness. Accordingly, we introduce the mutation brittleness metric that measures mutants' longevity and overflow between versions.



\begin{figure*}[!ht]
    \vspace{-1em}
     \centering
     \begin{subfigure}[t]{0.45\textwidth}
         \centering
         \includegraphics[trim={0 0 0 0},clip,width=\textwidth]{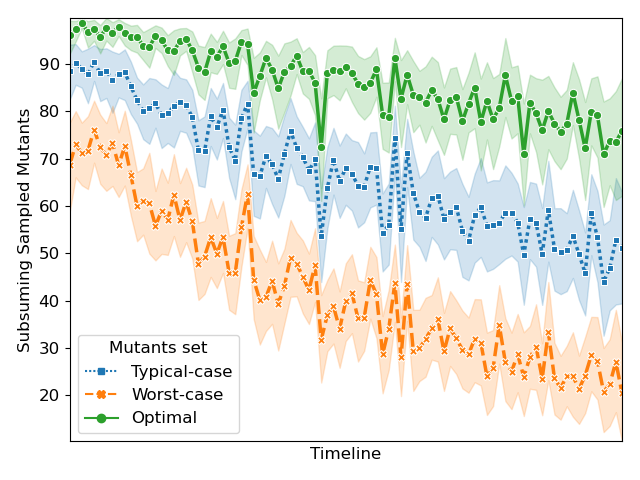}
         \vspace{-2em}
         \caption{\textbf{Long-Standing Subsuming mutants for different observed files throughout their studied \textit{history}. An optimal set of mutants - when selected - shows a long-standing prospect. In contrast, a worst-case set of mutants when selected shows brittleness - indicating obsolete test requirements. 
         A ‘special relationship’ between long-standing and subsuming mutants exists and indicates that  subsuming mutants in one version are probably subsuming in the following versions.
         }} 
         \label{fig:Distribution_of_sampled_mutants}
     \end{subfigure}
     \hfill
     \begin{subfigure}[t]{0.45\textwidth}
         \centering
         \includegraphics[width=\textwidth]{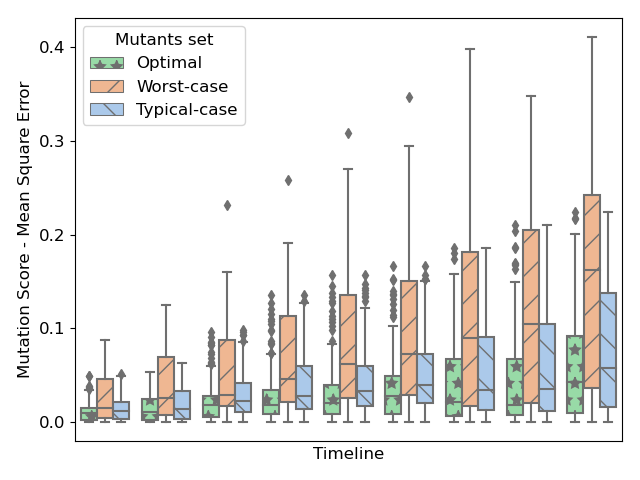}
         \vspace{-2em}
         \caption{\textbf{Mean Square Error of Mutation Score of initially selected and long-standing subsuming mutants. The optimal set of long-standing subsuming mutants demonstrates a capability to perform test assessments over time - preserving subsumption relationships - unlike the worst-case or typical sets which show higher MSE. An optimal set of long-standing mutant suites enjoy an order of magnitude longer relevance.}} 
         \label{fig:MS_MSE}
     \end{subfigure}
     \vspace{-2em}
    \caption{}
    \label{fig:Optimal_simulation}
    \vspace{-1.5em}
\end{figure*}

\section{Initial Evaluation and Early Results}

\subsubsection*{\textbf{Evaluation data}}

Table \ref{tab:Observed_history_of_files} shows our subject data. From 4 different well maintained Apache Commons projects, we extracted nine files with the longest history of change and their corresponding commits. It is important to emphasize that due to technical reasons (e.g., PIT mutation testing tool requires green test suite to run), the time gap between commits is rather "longer" than one commit. Nevertheless, this didn't stop us from mapping and observing long-standing mutants through the history of evolving systems.

\subsubsection*{\textbf{Mutation Brittleness}}

Figure \ref{fig:Distribution_of_long_living_mutants} depicts brittleness of mutants over time. In particular, it tells us how many mutants of a specific file exist through time, from their inception, on the same initial code elements, w.r.t., a code change has not altered mutants.
From the figure, we can observe the diversity of the longevity distribution. Moreover, for the file Lexer, the ratio of long-standing mutants is significant, over 80\% over observed time-points. On the contrary, we can see that the CsvPrinter file contains significant changes, and the ratio of the mutants degrade below 50\% after the first half of observed points and below 20\% in the second half of the observed history points. For other observed files, we see changes do not impact over 70\% of mutants in the first quartile of the timeline and between 40-60 \% for the rest of the time-points. These results demonstrate that mutants have a diverse lifetime over different evolution timelines, which suggests further investigation of whether mutants keep subsuming dynamic relationships over time and how mutant selection can affect test assessment. 


\subsubsection*{\textbf{Long-Standing Subsuming Mutants}} 

Figure \ref{fig:Optimal_simulation} demonstrates to what extent subsuming mutants convey their dynamic behaviour and how mutation selection can affect the test assessment capability of mutation testing over time. In particular, the figure depicts the scenario in which we select subsuming mutants at a certain point in time and observe the capacity in which they exist over time together with how well they can perform test assessment w.r.t., measuring mutation score.
We randomly sample 10-30\% of mutants (100 times to remove the threat of randomness; we choose these selection intervals as obviously selecting all mutants leads to traditional mutation testing) from each observed file and consider the file history length - \textit{the figures show aggregated results since each subject has different history length.} Figure \ref{fig:Distribution_of_sampled_mutants} illustrates the need for intelligent mutant selection as we can significantly distinguish between two sets, a) sets of mutants more optimal as they stand longer throughout observed history; hence longer enjoying mutant suites relevance and b) other sub-optimal sets that suffer relevance degradation, w.r.t., represent obsolete test requirements. Interestingly, optimal mutant selection promises continuous tracking of test effectiveness as the margin of degradation is $\approx$10\% on the ratio of selected mutants. In comparison, we observe a worst-case sets of mutants which indicate obsolete test requirements and typical arbitrary sets as if there was no other way to select, then we would end up with a random. To observe how capable those mutants are of affecting test assessment over time - keeping their subsumption relationships - we calculate the mean square error (MSE) of mutation score (MS) between the initially selected set and those long-standing mutant sets. In Figure \ref{fig:MS_MSE} we assess the difference in the MSE of MS between the optimal and suboptimal sets of long-standing mutants. We observe that the optimal set of long-standing subsuming mutants keeps MS high over time (low MSE $\approx$ 0.01\%-0.04\%), indicating a gradual loss in MS as the mutants stand longer in time, thus preserving mutant suite relevance. Accordingly, it is important to realize the potential in conveying knowledge of previously calculated dynamic relationships of mutants for at least 10x longer than a random selection.
In particular, by selecting the sub-optimal sets of mutants, the threat of not preserving the knowledge appears, w.r.t., mutants less capable of test assessment over time, suggesting their low priority (higher MSE $\approx$0.01\%-0.20\%). 

\vspace{-0.5em}
\section{Future Plans}
\vspace{-0.5em}
We believe that  long-standing mutants are an interesting category in their own right, and worthy of further research.
They have implications not only for mutation testing, but also beyond mutation testing.
In this section we set out future plans for further evaluation and investigation of the properties of long-standing mutants and their applications.


\textit{Implications regarding subsuming long-standing mutants:}
Despite showing that subsuming relationships can be preserved from version to version and that mutants' utility can be reused, we do not yet fully understand   {\em why} 
subsuming mutants tend to last longer than subsumed mutants.
A detailed study is needed to fully understand the subsumption and longevity drivers. 

\textit{Implications for mutation testing tools:}
Our results also have implications for the development of future mutation testing tools.
In particular, our results suggest the development of a robust mutant versioning system.
Existing tools \cite{ColesLHPV16,ChekamPT19,moran:mdroid} focus on the generation of mutants, but not sophisticated mutant versioning.
In future work, we need to investigate mutation testing tools that allow logging mutants' maturity, execution history, and fluctuation over time, supporting approaches that learn mutant behaviour and relating this to code changes.
Previous work on flaky mutant detection \cite{Shi0M19}, predictive modelling \cite{afzal:predictive-modeling-review} and hyper-heuristics \cite{mh:esem-keynote} (in particular that focused on mutation testing \cite{jzetal:predictive}) may form a good starting point for this research agenda.

\textit{Maximising long-standing mutant fault revelation:}
By focusing on long-standing mutants, we favour mutants that reside in relatively unchanging parts of the code.
There is a natural concern that this may, in turn, lead to us favouring test suites that do not tend to reveal faults in changing parts of the code.
Fortunately, the fact that a mutant lies in code region $A$  does not render it insensitive to bugs that lie in (lexically separate) code region $B$.
If there are transitive {\em dependencies} between $A$ on $B$ then we can expect high degrees of mutant coupling, and even subsumption between the two regions.
This suggests future work on identifying mutants that have high `transitive dependence reach' through their transitive dependencies, using techniques such as slicing \cite{dbetal:orbs-fse14} and chopping \cite{jackson:new}.

\textit{Implications of long-standing mutants beyond mutation testing research:}
The findings reported in this paper have implications beyond mutation testing to 
automated program repair \cite{legoues:cacm-survey,ametal:sapfix} 
and genetic improvement \cite{Petke:gisurvey,bill:evolving}.
It is often been argued that program repair is the inverse of mutation testing.
Instead of inserting faults, repair seeks to remove them.
Long-standing mutants are therefore also likely to find applications and implications in the field of program repair and genetic improvement research.
For example, it would be interesting to explore `long-standing' repairs as a counterpoint to long-standing mutants.
One might  reasonably conjecture that such repairs would remain relevant for longer than repairs in areas of code subject to high degrees of churn.
However, the empirical assessment of this phenomenon remains an open problem for future work.

\balance
\bibliographystyle{IEEEtran}
\bibliography{IEEEabrv,sample-base,mark-base}

\end{document}